\documentclass{ws-procs9x6}

\def\etal {{\it et al.}}

\newcommand{\squishlist}{
\begin{list}{$\bullet$}
{ \setlength{\itemsep}{-1pt}
\setlength{\parsep}{1pt}
\setlength{\topsep}{2pt}
\setlength{\partopsep}{0pt}
\setlength{\leftmargin}{2em}
\setlength{\labelwidth}{1.5em}
\setlength{\labelsep}{0.5em} } }

\newcommand{\squishend}{
\end{list} }

\def\aH{{\overline{\rm H}}{}}
\def\Ps{{\rm Ps}}

\begin{document}

\title{AEGIS AT CERN:\\
MEASURING ANTIHYDROGEN FALL}

\author{MARCO G.\ GIAMMARCHI}

\address{Istituto Nazionale di Fisica Nucleare\\
Via Celoria 16, Milano 20133, Italy\\
E-mail: marco.giammarchi@mi.infn.it}

\author{on behalf of the AEGIS Collaboration%
\footnote{
A.S.~Belov, G.~Bonomi, I.~Boscolo, R.S.~Brusa, V.M.~Byakov, L.~Cabaret, A.~Calloni, C.~Canali, C.~Carraro,
F.~Castelli, S.~Cialdi, D.~Comparat, G.~Consolati, N.~Djourelov, M.~Doser, G.~Drobychev, A.~Dudarev, A.~Dupasquier, D.~Fabris,
R.~Ferragut, G.~Ferrari, A.~Fischer, P.~Folegati, A.~Fontana, M.~Lunardon, M.G.~Giammarchi, S.N.~Gninenko, R.~Heyne, S.D.~Hogan, L.W.~Jorgensen,
A.~Kelleberbauer, D.~Krasnicky, V.~Lagomarsino, G.~Manuzio, S.~Mariazzi, V.A.~Matveev, F.~Merkt, S.~Moretto, C.~Morhard, G.~Nebbia,
P.~Nedelec, M.K.~Oberthaler, D.~Perini, S.~Pesente, V.~Petracek, M.~Prevedelli, I.Y.~Al-Qaradawi, F.~Quasso, C.~Riccardi, O.~Rohne,
A.~Rotondi, M.~Sacerdoti, H.~Sandaker, D.~Sillou, S.V.~Stepanov, H.H.~Stroke, G.~Testera, D.~Trezzi, A.V.~Turbabin, G.~Viesti,
F.~Villa, H.~Walters, U.~Warring, S.~Zavatarelli, A.~Zenoni, D.S.~Zvezhinskij.
}
}

\begin{abstract}
The main goal of the AEGIS experiment at the CERN Antiproton Decelerator is the test of 
fundamental laws such as the Weak Equivalence Principle (WEP) and CPT symmetry. 
In the first phase of AEGIS, a beam of antihydrogen will be formed whose fall in the gravitational 
field is measured in a Moir\'e deflectometer; this will constitute the first test of the WEP
with antimatter. 
\end{abstract}

\bodymatter

\section{Introduction}

The goal of AEGIS 
(Antimatter Experiment: Gravity, Interferometry, Spectroscopy)\cite{keller}, under construction at CERN, is the study
of fundamental physics with antimatter, namely the investigation of the Weak Equivalence Principle (WEP) and CPT symmetry. This
constitutes a test of the foundations of General Relativity and quantum field theory\cite{mavro,datatables}.

During the first phase of the experiment, the production of a antihydrogen ($\aH$) beam is foreseen to allow tests of the WEP by producing
antihydrogen through the charge-exchange reaction 
$(\Ps)^* \, \overline p \rightarrow \aH^* e^-$ and forming a $\aH$ beam whose
fall in the gravity field is measured with a Moir\'e deflectometer.

In a second phase of the experiment, $\aH$ will be laser--cooled and confined, to perform higher precision $g$ measurements and 
CPT tests.

\begin{figure}[!ht]
\center
\includegraphics[width=12cm,height=7cm,clip=true]{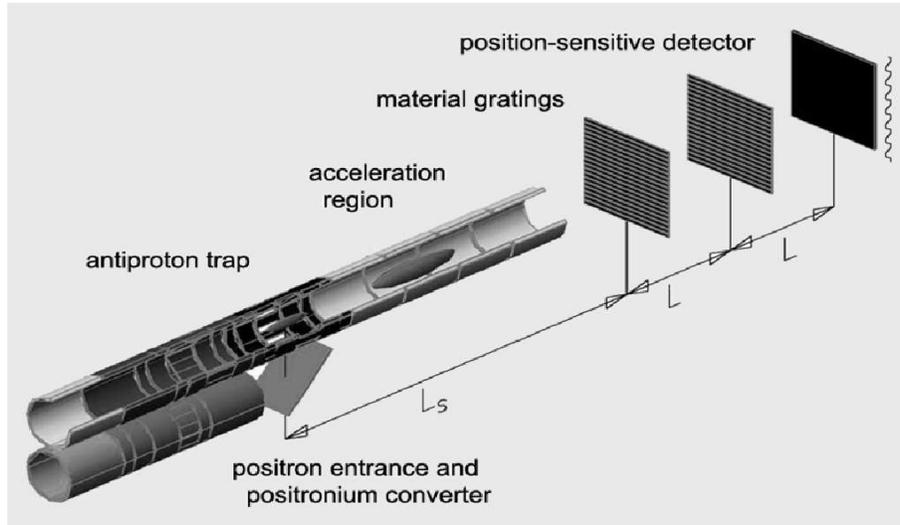}
\caption{Sketch of the central part of AEGIS. Two parallel Penning-Malmberg traps ($r = 8$ mm)
manipulate $\overline p$ and $e^+$ to form and accelerate $\aH$. They will be mounted inside a 100 mK
cryostat in a 1 T magnetic field. The upper trap is devoted to antiprotons. Cold $\overline p$ wait for Ps in the black 
region. The lower trap is devoted to positrons; they will be sent on to the converter to produce Ps. Laser pulses will 
excite the Ps to Rydberg states to form $\overline p$ in the black region. The acceleration region shows
the bunch of antihydrogen after the Stark acceleration. The two material gratings ($L_s=50$ cm, $L=30$ cm, transverse dimensions:
$20\times 20$ cm$^2$) and the detector are used for the $g$ measurement.}
\label{f:general}
\end{figure}

The best sensitivity for WEP tests made on matter systems come from rotating torsion balances\cite{schl} and from the Moon and Earth 
accelerations\cite{will}; they are both in the range of $\sim 10^{-13}$. On the other hand, there have been no  
direct measurements of the gravitational acceleration of antimatter. 

\section{The production of the antihydrogen beam}

The measurement of $g$ is performed using the gravitational fall of a 
beam of antihydrogen, whose production involves the following steps.

\squishlist

\item Production of positrons in a Surko-type source and accumulator

\item Accumulation and cooling of the antiprotons 

\item Production of Ps by positron-bombardment of a converter

\item Laser excitation of the Ps to a $n \simeq 20-25$ Rydberg state

\item Production of $\aH$ by means of the reaction $(\Ps)^* \, \overline p \rightarrow \aH^* e^-$

\item Formation of a $\aH$ beam by Stark acceleration

\item Measurement of $g$ in a Moir\'e deflectometer

\squishend

Figure \ref{f:general} shows a schematics of the core of the apparatus, whose antiproton (Penning-Malberg) trap is immersed in a 1 T magnetic field
for charged particle confinement.

After antiproton cooling (requiring about 300 s), the positron bunch is fired at the converter to produce Ps for the charge exchange reaction. 
The production process is therefore a pulsed production triggered by the $e^+$ hitting the converter. The produced Ps gets laser excited before 
intersecting the $\overline p$ cloud. The $\aH$ production process takes about a microsecond; after that, the antiatoms 
are accelerated towards the deflectometer by means of a Stark acceleration technique.
The $\overline p$ beam fall will be measured with a Moir\'e deflectometer equipped with a position sensitive microstrip detector.

Antiprotons are coming from the CERN Antiproton Decelerator (AD), delivering about $3\times 10^7$ particles every 100 s at 6 MeV. After an energy 
degrader, antiprotons are caught in a 3 T magnetic field region (not shown in Fig.\ \ref{f:general}) at a temperature of 4 K and further cooled 
down to 100 mK (a velocity of $\sim$ 50 m/s). This low temperature is achieved by resistive cooling or sympathetic cooling with laser cooled 
osmium ions\cite{bauer}. 

\begin{figure}[!ht]
\center
\includegraphics[width=10cm,height=6cm,clip=true]{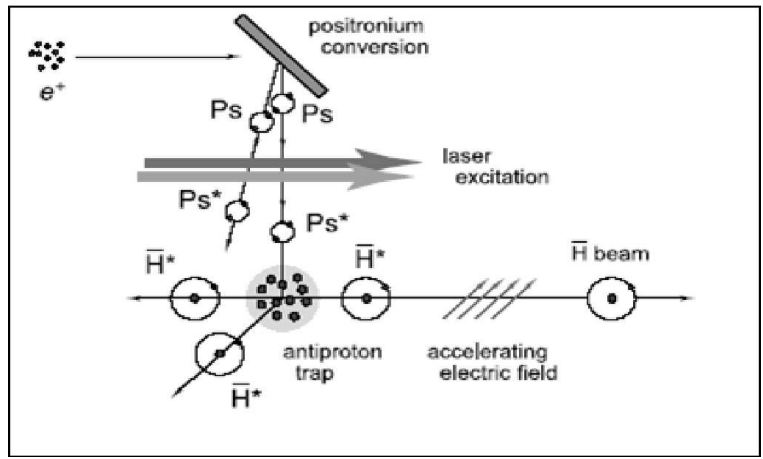}
\caption{Schematics of the AEGIS antihydrogen production and acceleration process. The positron bunch lasts about 20 ns while the laser pulse
is 10 ns long.}
\label{smaller}
\end{figure}

\subsection{Positronium formation and excitation}

Positronium will be formed by an $e^+$ bunch hitting a converter. Positrons will be produced by a Surko-type accumulator 
delivering a 20 ns bunch of about $10^8$ particles with a few keV energy.

After Ps has been formed in the bulk of the converter material, ortho-Ps can be reemitted out of the target (Fig.\ \ref{smaller}). 
experiments have shown that in suitable converter materials, reemitted Ps in the amount of about 50\% of the impinging $e^+$ can be 
obtained\cite{lizkay,brusa,ferragut}.  

The excitation of Ps to Rydberg states is necessary since the cross section of the reaction $(\Ps)^* \, \overline p \rightarrow \aH^* e^-$ 
has a strong dependence on the principal quantum number of the Ps, $\sigma \sim a_0 n^4$ ($a_0$ being the Bohr radius). Positronium
will be laser excited to a high-$n$ (Rydberg) state using a $1\rightarrow 3\rightarrow n$ excitation process, where $n=20-25$\cite{castelli}.

\subsection{Antihydrogen formation and acceleration}

Taking into account the production cross section, the geometry of the system and the number of $\overline p$ and excited Ps atoms, about  
500 antihydrogen atoms will be formed per positron shot on the converter. Since this process follows
the accumulation of $e^+$ in the positron accumulator and the accumulation and cooling of $\overline p$ in the catching region (300 s), 
the $\aH$ averaged production rate will be of a few Hz. The low temperature of the antiproton cloud allows the production of antihydrogen 
with an energy of 100 mK, or a thermal speed of 50 m/s (similar to the antiproton energy). 

The produced $\aH$ will be Stark-accelerated along the beam axis using the technique recently demonstrated for hydrogen\cite{vliegen2}. 
The $\aH$ atoms will arrive at the beginning of the deflectometer with an axial velocity of 200-500 m/s and a radial velocity spread 
of $\sim$ 50 m/s. 

\section{The gravity measurement}

\begin{figure}[!ht]
\center
\includegraphics[width=8cm,height=4.7cm,clip=true]{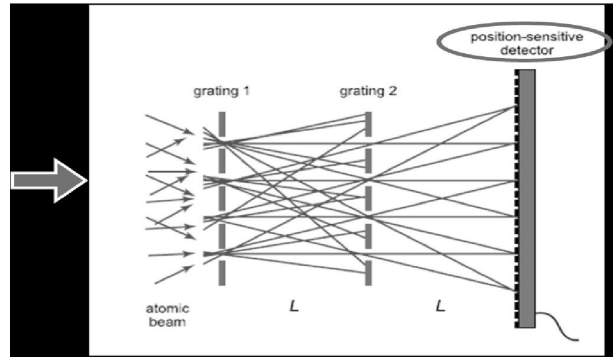}
\caption{Principle of the Moir\`e deflectometer and the detector for AEGIS. $L$ is 30 cm.}
\label{defle}
\end{figure}

Measuring $g$ with a flight path of 60 cm and a velocity $v=500$ m/s involves measuring a displacement of $\simeq 20$ $\mu$m against an 
8 mm beam spot. This will be done with a classical Moir\'e deflectometer; the device\cite{markus} consists of three equally spaced and parallel 
material gratings (Fig.\ \ref{defle}). The last plane will be a position-sensitive microstrip detector to register the time and impact point of 
$\aH$ atoms. 

As the atomic beam passes through the gratings the first two planes select specific propagation directions creating on the third plane a density 
modulation repeating itself at positions that are integer multiples of the distance between the first two gratings.
This technique, 
originally proposed in Ref.\ \refcite{kafri}, can be effectively applied to the case of inertial sensing (and gravity measurements) as 
discussed in Ref.\ \refcite{markus}. 

The modulation intensity pattern will be shifted by a quantity $\delta$ that depends on 
the transit time $T$, the period $a$ of the grating and $g$: $\delta = gT^2/a$. The gravity constant $g$ will be measured by fitting
this distribution. Taking into account several possible sources of errors, a final 1\% resolution on $g$ can be achieved 
by launching $10^5$ antihydrogen atoms towards the deflectometer. This amounts to a few months of data taking.


\begin{thebibliography}{xx}

\bibitem{keller}
A.\ Kellerbauer \etal, 
Nucl.\ Instr.\ Meth.\ B {\bf 266}, 351 (2008).

\bibitem{mavro}
N.E.\ Mavromatos, 
in A.\ Hirtl, J.\ Marton, E.\ Widmann, and J.\ Zmeskal, eds.,
{\it International Conference on Exotic Atoms and Related Topics}, 
Austrian Academy of Sciences, Vienna, 2006. 

\bibitem{datatables}
{\it Data Tables for Lorentz and CPT Violation,}
2010 edition,
V.A.\ Kosteleck\'y and N.\ Russell,
arXiv:0801.0287v3.

\bibitem{schl} 
S.\ Schlamminger \etal,
Phys.\ Rev.\ Lett.\ {\bf 100}, 041101 (2008).

\bibitem{will} 
J.G.\ Will \etal,
Phys.\ Rev.\ Lett.\ {\bf 93}, 261101 (2004).

\bibitem{bauer} 
U.\ Warring \etal,
Phys.\ Rev.\ Lett.\ {\bf 102}, 043001 (2009).

\bibitem{lizkay} 
L.\ Lizkay \etal,
Appl.\ Phys.\ Lett.\ {\bf 92}, 063114 (2008).

\bibitem{brusa} 
S.\ Mariazzi \etal,
Phys.\ Rev.\ B {\bf 68}, 085428 (2008).

\bibitem{ferragut}
R.\ Ferragut \etal,
submitted to Can.\ J.\ Phys.

\bibitem{castelli}
F.\ Castelli \etal,
Phys.\ Rev.\ A {\bf 78}, 052512 (2008).

\bibitem{vliegen2} 
E.\ Vliegen \etal,
Phys.\ Rev.\ A {\bf 76}, 023405 (2007).

\bibitem{markus} 
M.K.\ Oberthaler \etal,
Phys.\ Rev.\ A {\bf 54}, 3165 (1996). 

\bibitem{kafri}
O.\ Kafri,
Opt.\ Lett.\ {\bf 5}, 555 (1980).

\end{thebibliography}
\end{document}